\documentclass[aps,prb,showpacs,twocolumn,groupedaddress]{revtex4-1}
\usepackage{times,xspace}
\usepackage{graphicx,graphics,color,epsfig}
\usepackage{amsmath}
\usepackage{amssymb}
\usepackage{amsfonts}
\usepackage{amsfonts}
\usepackage{epstopdf}
\usepackage{bm}
\usepackage{hyperref}
\usepackage{color}
\usepackage{float}
\usepackage[normalem]{ulem}

\def\be{\begin{equation}}
\def\ee{\end{equation}}
\def\ba{\begin{eqnarray}}
\def\ea{\end{eqnarray}}

\def\moire{Moir\'e}

\begin{document}

\title{Wannier pairs in the superconducting twisted bilayer graphene and related systems}

\author{Sujay Ray$^1$, Jeil Jung$^2$, and Tanmoy Das$^1$}
\email[]{tnmydas@gmail.com}
\affiliation{$^1$Department of Physics, Indian Institute of Science, Bangalore, India - 560012.
$^2$Department of Physics, University of Seoul, Seoul 02504, Korea.}

\date{\today}

\begin{abstract}
Unconventional superconductivity often arises from Cooper pairing between neighboring atomic sites, stipulating a characteristic pairing symmetry in the reciprocal space. The twisted bilayer graphene (TBG) presents a new setting where superconductivity emerges on the flat bands whose Wannier wavefunctions spread over many graphene unit cells, forming the so-called Moir\'e pattern. To unravel how Wannier states form Cooper pairs, we study the interplay between electronic, structural, and pairing instabilities in TBG. For comparisons, we also study graphene on boron-nitride (GBN) possessing a different Moir\'e pattern, and single-layer graphene (SLG) without a Moir\'e pattern. For all cases, we compute the pairing eigenvalues and eigenfunctions by solving a linearized superconducting gap equation, where the spin-fluctuation mediated pairing potential is evaluated from materials specific tight-binding band structures. We find an extended $s$-wave as the leading pairing symmetry in TBG, in which the nearest-neighbor Wannier sites form Cooper pairs with same phase. In contrast, GBN assumes a $p+ip$-wave pairing between nearest-neighbor Wannier states with odd-parity phase, while SLG has the $d+id$-wave symmetry for inter-sublattice pairing with even-parity phase. Moreover, while $p+ip$, and $d+id$ pairings are chiral, and nodeless, but the extended $s$-wave channel possesses accidental {\it nodes}. The nodal pairing symmetry makes it easily distinguishable via power-law dependencies in thermodynamical entities, in addition to their direct visualization via spectroscopies.
\end{abstract}


\maketitle

\section{Introduction}
\begin{figure}[t]
\includegraphics[width=85mm,scale=1.]{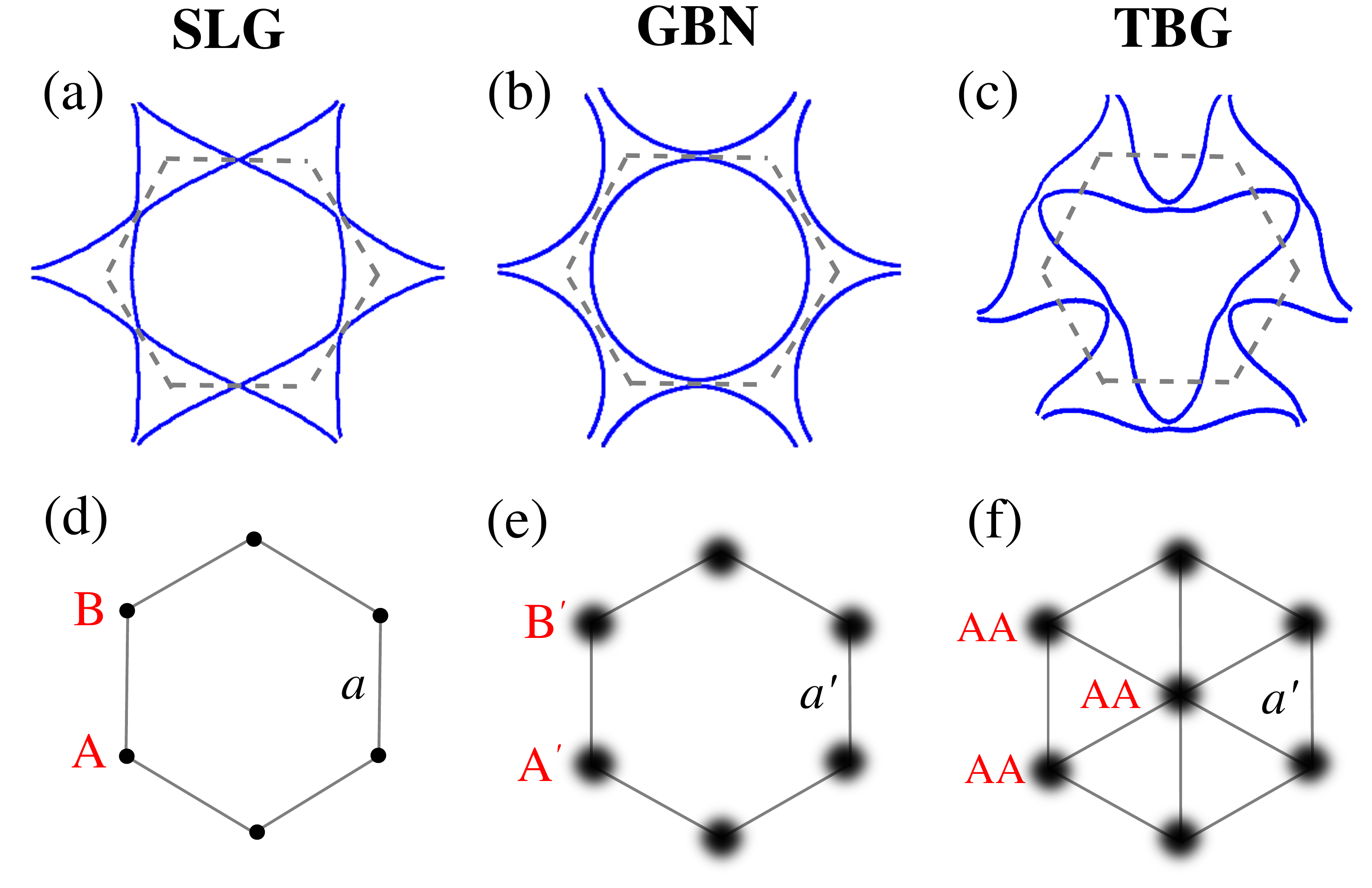}
\caption{(a-c) Computed FSs of SLG, GBN, and TBG, respectively at their corresponding VHS energies (dashed line depicts the 1st BZ). (d-f) Corresponding positions of the Wannier states of the VHS/flat band in the direct lattice. For SLG [(a) and (d)], the VHS's Wannier states are localized on the `A', and `B' sublattices in the primitive unit cell. In GBN [(b) and (e)], the Wannier states are localized on the corners of the hexagonal \moire-supercell. In TBG [(c) and (f)], the Wannier states show a fully formed triangular lattice at the flat band for {\it each valley}, where `A' sublattices of the original two graphene lattices merge on top of each other (defined as `AA' site). The Wannier states in both \moire-lattices spread over several graphene unit cell. $a=2.46\AA$ is the graphene's lattice constant, while $a'$ is the \moire~lattice constant.
}
\label{fig1}
\end{figure}

Strongly correlated quantum phases and superconductivity have long been predicted in single-layer graphene (SLG) at the van-Hove singularity (VHS).\cite{SLGVHS} However, their experimental realization has so far remained elusive. Recently, both correlated insulating gap\cite{TBGMott} and superconductivity\cite{TBGSC} have been observed in a twisted bilayer graphene (TBG) at a narrow range of twist angles, namely the `magic' angles $\sim 1^{o}$. In this region, the single-particle density of states (DOS) acquires a sharp peak near the Fermi level, with an effective bandwidth reducing to $\sim 5$ meV.\cite{MacDonald,JJ} The emergence of this flat band is intrinsic to the physics of \moire~pattern, formed in TBG as well as in graphene on hexagonal Boron Nitride (GBN).\cite{MacDonald,JJ,WW_JJ} The \moire~superlattice produces `cloned' Dirac cones at the \moire~zone boundaries, in addition to the primary Dirac cone at the \moire~zone center. The band dispersion between the primary and cloned Dirac cones pass through saddle-points or VHSs, and hence yields a flat band. It is tempting to assume that the `magic' angle creates a similar VHS-like state as in SLG and/or GBN, and thus the predicted correlated physics of SLG/GBN are also at play in TBG. However, a closer look at the electronic instabilities at the VHS and their characteristic localizations into unique Wannier states in the direct lattice reveals stark differences between them (see Fig.~\ref{fig1}). This leads to an essential question: How do such Wannier states, enveloping many graphene unit cells, condensate into Cooper pairs?

The relationships between the $k$-space electronic structure and direct lattice Wannier states of the SLG, GBN, and TBG are delineated in Fig.~\ref{fig1}. The effective bandwidth of the VHS/flat band decreases from $\sim$1 eV in SLG to $\sim$100 meV in GBN to $\sim$3-5 meV in TBG, making the latter more prone to correlation. Fermi surface (FS) of SLG, GBN, and TBG are compared in Fig.~\ref{fig1} at their corresponding VHS position. The FS of SLG is most flat (producing large nesting), while that for GBN is most circular (weak nesting), and TBG lies in between. In addition, we observe a systematic transition from six-fold to three-fold rotational symmetry in going from SLG to GBN to TBG, rearranging the corresponding Wannier states accordingly in the direct lattice. The three-fold symmetic FS of TBG is for a given valley band, while the other valley band has the complementary three-fold symmetry so that the FS becomes six-fold symmetric when both valley bands are included.\cite{hexagonallattice,hexagonallattice1} This {\it three-fold} symmetric FS makes TBG distinct from other hexagonal\cite{SLGVHS} and triangular lattices\cite{NaCoO2} with six-fold symmetric FS and plays an important role in stabilizing a distinct pairing symmetry here.

One of the most striking differences  emerges when we investigate the corresponding Wannier states of individual flat band in the direct lattice, see Fig.~\ref{fig1}(lower panel). In the flat region of the VHS in SLG near the {\bf K}-point, the states are localized on the `A' sublattices, while the states near the {\bf K$^{\prime}$} point are localized on the `B'-sublattice and vice versa. In GBN and TBG, the situation changes drastically due to \moire-supercell formation. In the low-energy model of the GBN \moire-lattice, the band structure can be described by that of a SLG under an effective supercell potential due to BN substrate with the supercell periodicity being much larger than the graphene unit cell. The corresponding Wannier states are maximally localized only on the  corners of the hexagonal \moire-supercell (enclosing several `A' and `B' sublattices of the original graphene unit cell),\cite{JJ} see Fig.~\ref{fig1}(e). On the contrary, in TBG the Wannier states of a given valley band are maximally localized on the `AA' lattice sites (where `A' sublattices of both graphene layers become aligned on top of each other) at all \moire-supercell corners, as well as at the center, forming a full triangular symmetry,\cite{TBGMott,TBGSC,macdonaldcorrelations} see Fig.~\ref{fig1}(f). The other valley band is also localized on the same `AA' sites, forming an unit cell with {\it two Wannier orbitals per site}, with different orbitals possessing complementary rotational symmetry.\cite{MacDonald,JJ,hexagonallattice,hexagonallattice1} 

We perform the pairing symmetry calculation using materials specific, multiband Hubbard model. Hubbard model has a SC solution arising from the repulsive many-body pairing interaction which mediates unconventional, sign-reversal pairing symmetry.\cite{SCrepulsive} Such a mechanism, often known as spin-fluctuation mediated unconventional superconductivity, basically depends on strong FS nesting instability at a preferred wavevector, say ${\bf Q}$. The nesting can promote a SC solution with a momentum-dependent pairing symmetry $\Delta_{\bf k}$ such that the pairing symmetry changes sign on the FS as: ${\rm sgn}[\Delta_{\bf k}]=-{\rm sgn}[\Delta_{\bf k+Q}]$. This sign reversal is required to compensate for the positive (repulsive) pairing potential. This theory of spin-fluctuation driven superconductivity consistently links between the observed pairing symmetry and FS topology in many different unconventional superconductors.\cite{spinfluc,SCpnictides,SCHF,SCorganics} A ${\bf k}$-dependent pairing symmetry incipiently requires that pairing occurs between different atomic sites in the direct lattice. In what follows, the characteristic momentum structure of the pairing symmetry is intimately related to the underlying pairing mechanism, FS topology, and its contributing Wannier sites. 

For each material, we obtain the non-interacting, low-energy band structures by tight-binding model in the unit cell or \moire~cell, as appropriate. 
Next we solve the pairing eigenvalue (SC coupling constant) and eigenfunction (pairing symmetry) solution of the linearized Eliashberg equation, where the pairing potential stems from many-body spin- and charge fluctuations.\cite{spinfluc,SCpnictides,SCHF,SCorganics} The obtained eigenfunction for the largest eigenvalue gives the pairing symmetry in the momentum space. We obtain the real-space mapping of the pairing symmetry by inverse Fourier transform. This illuminates the Cooper pairs between the nearest neighbor Wannier orbitals with corresponding phase factor. 

In SLG, the computed pairing eigenfunction agrees with a $d+id$-wave symmetry, which arises from inter-sublattice pairing between the `A' and `B' Wannier sites in a hexagonal primitive lattice. In GBN, the pairing solution changes to a $p+ip$-symmetry where the inter-sublattice pairing occurs between the nearest neighbor (NN) Wannier orbitals with odd-parity phases. On the other hand, in TBG, we find an extended $s$- pairing with even parity phases between the same Wannier orbitals in NN sites. Note that the extended $s$-wave solution can produce accidental nodes when the FS is large near the VHS doping.

The rest of the manuscript is organized as follows. In Sec.~\ref{Sec:Cal} we discuss  the computational details for the electronic structure calculations, and pairing eigenvalue calculations. All results are presented in discussed in Sec.~\ref{Sec:Results}. Finally we conclude in Sec.~\ref{Sec:Discussion}.

\section{Theory}\label{Sec:Cal}

\subsection{Electronic structure and FS nestings}\label{Sec:Band}

For SLG, we use a typical two band tight-binding (TB) model as presented in the literature.\cite{CastroNetoRMP,JJWannier} For the \moire-lattices in GBN and TBG, we directly use the TB model presented in Refs.~\onlinecite{MacDonald}\onlinecite{JJ}. As we are interested in the low-energy properties, we downfold all the bands into an effective low-energy six band model.\cite{TB_technicaldetails} Details of each band structure parameterization are given in Appendix~\ref{AppA}. In Fig.~\ref{fig1} (top panel), we show computed FS topology for the three systems under study with the chemical potential placed at the VHS/flat band. In the corresponding lower-panel of Fig.~\ref{fig1}, we show the Wannier states for the Fermi momenta on the flat band.

To estimate the FS nesting features, and the corresponding pairing potential, we compute the multiband Lindhard susceptibility $\chi_{\alpha\beta}({\bf q},\omega)$:
\begin{eqnarray}
\chi_{\alpha\beta}(\omega,{\bf q})=-\sum_{\bf k}F_{\nu\nu'}^{\alpha\beta}({\bf k,q})\frac{f(\epsilon_{\bf k}^{\nu})-f(\epsilon_{\bf k+q}^{\nu'})}{\omega+i\delta-\epsilon_{\bf k}^{\nu}+\epsilon_{\bf k+q}^{\nu'}},
\label{Eq:Bare}
\end{eqnarray}
where $\xi^{\nu}_{\bf k}$ is the $\nu^{\rm th}$ band, and  $f(\xi_{\bf k}^{\nu})$ is the corresponding fermion occupation number. $\alpha$, $\beta$ give the orbital indices, and ${\bf q}$ and $\omega$ are the momentum and frequency transfer, respectively.  $F_{\bf k,q}$ as form factor arising from the eigenvectors as 
\begin{equation}
F_{\nu\nu'}^{\alpha\beta}({\bf k,q}) = u_{\alpha}^{\nu\dag}({\bf k})u_{\beta}^{\nu}({\bf k})u_{\beta}^{\nu'\dag}({\bf k+q})u_{\alpha}^{\nu}({\bf k+q}),
\end{equation}
where $u_{\alpha}^{\nu}$ represents the eigenvector for the $\nu^{\rm th}$-band projected to the $\alpha^{\rm th}$ basis (Wannier orbitals). We evaluated the form-factor numerically.\cite{footform}

\begin{figure}[t]
\includegraphics[width=85mm,scale=1.]{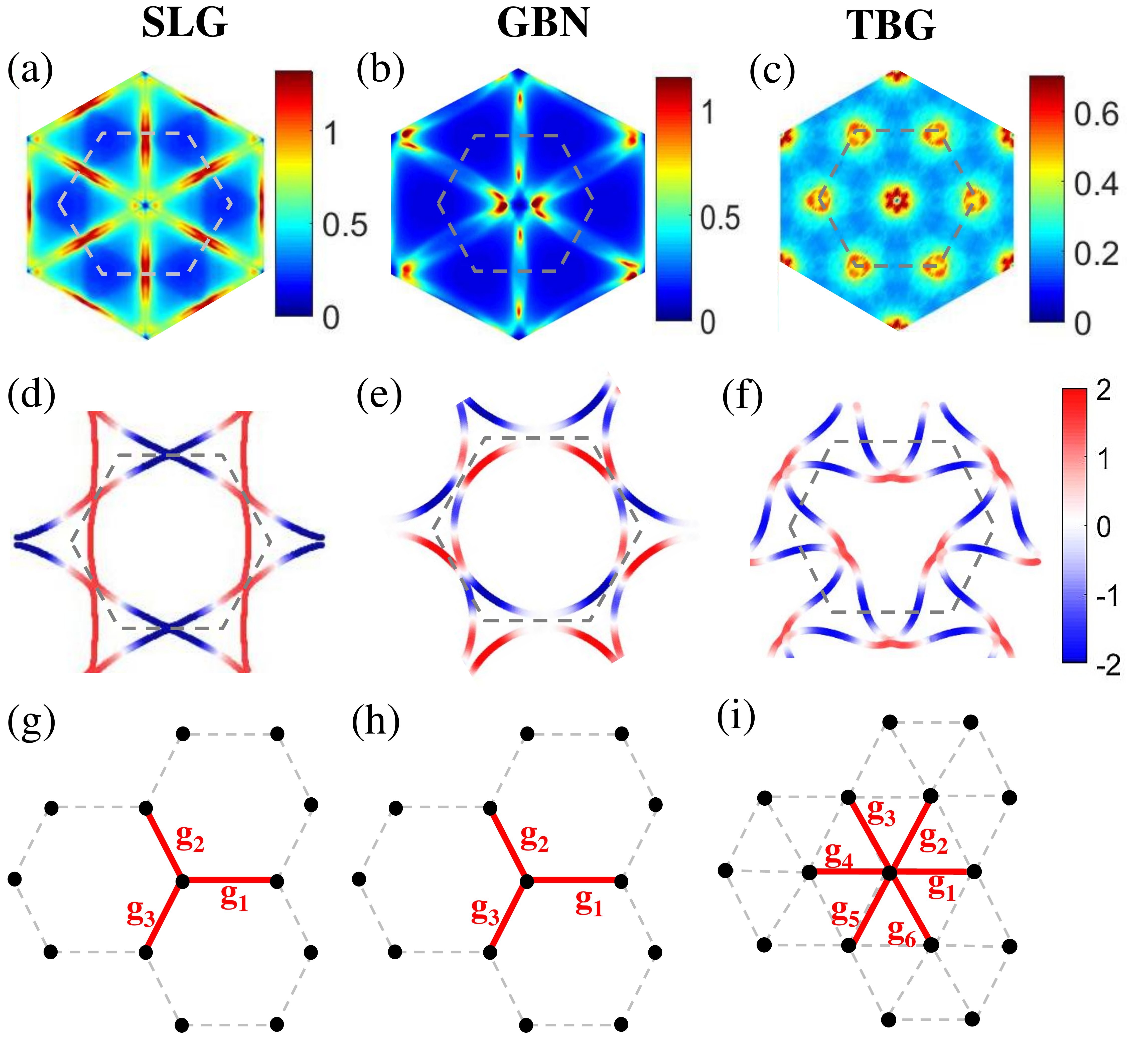}
\caption{(a-c) Spin susceptibility within RPA approximation for (a) SLG, (b) GBN and (c) TBG. (d-f) Computed pairing eigenfunctions for the highest eigenvalue of Eq.~\eqref{eq:lambda} for (d) SLG, (e) GBN and (f) TBG at their VHS dopings are plotted on the FS in a blue (negative) to white (nodes) to red (positive) colormap. The pairing structure is consistent with a $d+id$-wave and $p+ip$-wave symmetry in SLG, and GBN, respectively. On the other hand for TBG in (f) we find a rotationally invariant extended $s$-wave symmetry. (g-i) The real space picture of the pairing for (g) SLG, (h) GBN and (i) TBG systems. $g_{j}$ denote the pairing strength between nearest sites which is obtained from Fourier transformation of corresponding pairing functions [Eq.~\eqref{InvFT}].
}
\label{fig2}
\end{figure}

We present the 2D profile of the susceptibility (total $\chi=\sum_{\alpha\beta}\chi_{\alpha\beta}$) for $\omega\rightarrow 0$ in Fig.~\ref{fig2} (top panel) for all three systems. We find stark differences in the nesting features. In SLG, the FS is extremely flat, causing paramount FS nesting at ${\bf Q}\sim (2/3, 1/3)$ r.l.u., and its equivalent points. The nesting is considerably weak in GBN since here the FS is quite circular, with some residual nesting occurring at small wavevectors. For TBG, the nesting is strong at ${\bf Q}\sim (1/3, 0)$ r.l.u.. Such a FS nesting drives translation symmetry breaking into various density-wave orders in the particle-hole channels and/or unconventional pairing instability. The FS nesting driven superconductivity stabilizes a characteristic symmetry which changes sign on the FS. 

\subsection{Pairing symmetry calculations}\label{Sec:Pair}
Next we compute the pairing symmetry and pairing strength arising from the density-density fluctuations. It should be noted that although the bandwidth is lower near the magic angles, the FS becomes large due to VHS. This enhances screening, and hence the effective Coulomb interaction is reduced.\cite{ADas} The largest insulating gap obtained near half-filling in TBG is $\sim$0.3 meV $<$ bandwidth, rendering an effective weak or intermediate coupling regime for correlation. For such a correlation strength, the many-body density-density (spin and charge) correlation functions are computed from multiband Hubbard model. Since we restrict our doping range to only within individual flat bands, the corresponding intra-band Hubbard $U$ dominate the correlation spectrum. The multiband Hubbard interaction reads as
\begin{eqnarray}
H_{int} = \frac{1}{\Omega_{\rm BZ}}\sum_{\alpha\alpha'} U_{\alpha\beta}\sum_{{\bf q},\sigma\sigma'}n_ {\alpha\sigma}({\bf q})n_{\beta\sigma'}(-{\bf q}),
\label{HubbardU}
\end{eqnarray}
where $n_{\alpha\sigma}({\bf q})$ is the density operator for the $\alpha^{\rm th}$-band with $\sigma=\uparrow,\downarrow$ spins, and $U_{\alpha\beta}$ is the Hubbard $U$ between the two bands. Based on this Hubbard model, we compute pairing potential from the bubble and ladder diagrams to obtain for singlet and triplet channels as\cite{spinfluc,SCpnictides,SCHF,SCorganics}
\begin{eqnarray}
\tilde{\Gamma}^{\rm s}({\bf q})&=& \frac{1}{2}{\rm Re}\big[3{\tilde U}_s{\tilde \chi}^s({\bf q}){\tilde U}_s - {\tilde U}^c{\tilde \chi}^c({\bf q}){\tilde U}_c + {\tilde U}_s+{\tilde U}_c\big],
\label{singlet}\\
\tilde{\Gamma}^{\rm t}({\bf q})&=&-\frac{1}{2}{\rm Re}\big[{\tilde U}_s{\tilde \chi}^s({\bf q}){\tilde U}_s + {\tilde U}_c{\tilde \chi}^c({\bf q}){\tilde U}_c - {\tilde U}_s-{\tilde U}_c\big].
\label{triplet}
\end{eqnarray}
Here we introduce `tilde' to symbolize a quantity to be a matrix of dimension $N\times N$, with $N$ being the total number of bands. Superscript `s', and `c' denote many-body  spin and charge susceptibilities $\tilde{\chi}^{s/c}({\bf q})$ matrix whose components are defined as
\begin{eqnarray}
\chi_{\alpha\beta}^{\rm s/c}=\chi_{\alpha\beta}(1 \mp U^{\rm s/c}_{\alpha\beta}\chi_{\alpha\beta})^{-1}.
\end{eqnarray}
Here $\chi_{\alpha\beta}$ is the bare susceptibility defined in Eq.~\eqref{Eq:Bare} above. 
The many-body susceptibilities are obtained within the random phase approximation (RPA). $U_{s/c}$ are the Hubbard $U$ matrix for spin-flip and non spin-flip interactions, respectively (Eq.~\eqref{HubbardU}). Here $U^{s/c}_{\alpha\beta}=U$ for $\alpha=\beta$ and $U^{s/c}_{\alpha\beta}=V$ for $\alpha \ne \beta$. $U$ differs in different systems.\cite{U} Clearly, larger $U$ increases (decreases) spin (charge) susceptibility. Essentially in moderate coupling regime, spin-fluctuation dominates while charge sector acts as pair-breaker for the spin-singlet pairing ($\Gamma^{\rm s}$ in Eq.~\eqref{singlet}). 

A triplet pairing channel  $\Gamma^{\rm t}$ increases when the onsite interaction dominates over spin and charge fluctuations, as in the case of GBN (see below). In both singlet and triplet cases, it is evident that the pairing potentials have strong peaks at the momenta where the underlying susceptibility itself obtains peaks, i.e., pairing potentials $\Gamma^{\rm s/t}({\bf q})$ also diverge at the FS nesting wavevectors, and hence stabilize a characteristic pairing symmetry in a given system.

Based on the above pairing potential, we solve the linearized multiband SC gap equation, which is the pairing eigenvalue equation, as given by (see Appendix~\ref{AppB} for details)
\begin{equation}
\lambda_{\nu} g_{\nu}({\bf k}_{\alpha})=-\frac{1}{\Omega_{\rm FS}}\sum_{\beta,{\bf k}_{\beta}^\prime}
\Gamma^{\nu}_{\alpha\beta}({\bf k}_{\alpha}-{\bf k}_{\beta}^{\prime})g_{\nu}({\bf k}_{\beta}^{\prime}),
\label{eq:lambda}
\end{equation}
where ${\bf k}_{\alpha}$ is the Fermi momentum for the $\alpha^{\rm th}$ band. The eigenvalue calculation is performed over the entire 2D FS to estimate the dominant eigenvalue $\lambda$ (which measures the SC coupling constant), and the corresponding eigenvector gives the leading pairing symmetry $g({\bf k})$. The same eigenvalue equation is solved for both singlet ($\nu\equiv$ s) and triplet ($\nu\equiv$ t) channels. Since the pairing potentials $\Gamma^{\rm s/t}$ scale with the Hubbard $U$, SC coupling constant $\lambda$ also increases with increasing $U$. Within the first-order approximation, the pairing symmetry $g({\bf k})$ does not scale with $U$ (in the weak to moderate coupling regime). Therefore, our general conclusions about the pairing symmetry, and the phase diagram are dictated by the nesting strength, and remain valid for different values of $U$ in this coupling regime.

For a repulsive interaction $\Gamma^{\nu}>0$, according to Eq.\eqref{eq:lambda}, a positive eigenvalue $\lambda$ can commence with the corresponding eigenfunction $g({\bf k})$ changing sign as ${\rm sgn}[g({\bf k})]=- {\rm sgn}[g({\bf k'})]$ mediated by strong peak(s) in $\Gamma^{\nu}$ at ${\bf Q}=\mathbf{k}-\mathbf{k^{'}}$. Looking into the origin of $\Gamma^{\nu}$ in Eqs.~\eqref{singlet}, \eqref{triplet}, we notice that $\Gamma^{\nu}$ inherits strong peaks from that in $\chi^{s/c}$, which is directly linked to the FS nesting feature embedded in $\chi$. 

\section{Results}\label{Sec:Results}
Here we discuss our results of the pairing eigenstates for three systems under considerations at their VHS dopings. The computed results of $g({\bf k})$ for the largest eigenvalue of Eq.~\eqref{eq:lambda} are shown in the middle panel in Fig.~\ref{fig2}. The momentum space symmetry of the eigenfunction $g({\bf k})$ is obtained by comparing with the orbital symmetry of the spherical harmonics. After that we inverse Fourier transform the $g({\bf k})$ to the unit cell/Moire superlcell as
\begin{eqnarray}
g_j=\frac{1}{\Omega_{\rm BZ}}\sum_{\bf k}g({\bf k})e^{-i({\bf k}.{\bm \delta}_{j}-\phi_{\bf k})},
\label{InvFT}
\end{eqnarray}
where $g_j$ gives the pairing amplitude between two Wannier sites separated by a distance ${\bf \delta}_j$, see Fig.~\ref{fig2}(g-i). $\phi_{\bf k}={\rm Arg}[\sum_j e^{-i{\bf k}.{\bm \delta}_{j}}]$ is an additional phase factor arising in the hexagonal lattice possessing two Wannier basis per unit cell.\cite{SLG_pairing} We discuss below each system separately. 


\subsection{SLG}
For SLG, numerous calculations predicted that an exotic $d_{x^2-y^2}+id_{xy}$ ($d+id$) - wave symmetry is the dominant pairing channel, constrained by the FS nesting at the VHS.\cite{SLGVHS} We also find here that the two highest eigenvalues are the same with $\lambda=0.26$ with the corresponding degenerate eigenfunctions being 
\begin{eqnarray}
g^{d_{x^2-y^2}}({\bf k})&=&\cos\left(k_{y}-\phi_{\bf k}\right)+\cos\left(\frac{k_{y}}{2}+\phi_{\bf k}\right)\cos\left(\frac{\sqrt{3}k_{x}}{2}\right),\nonumber\\
g^{d_{xy}}({\bf k})&=&\sin\left(\frac{k_{y}}{2}+\phi_{\bf k}\right)\sin\left(\frac{\sqrt{3}k_{x}}{2}\right).
\label{gkSLG1}
\end{eqnarray}
These two eigenfunctions, respectively, represent $d_{x^2-y^2}$ and $d_{xy}$ symmetries in the hexagonal BZ. Because of the degeneracy, the $E_{2g}$ irreducible representation allows a complex mixing between them which is called the $d+id$-pairing symmetry in SLG.\cite{KL_doped_graphene} (We repeat the calculation with different $U$, the absolute value of the eigenvalue changes, but the eigenfunctions remain the same). In Fig.~\ref{fig2}(d) we show the $d_{x^2-y^2}$ eigenfunction, overlaid on the corresponding FS a color-gradient scale. Using Eq.~\eqref{InvFT}, we obtain pairing amplitude between three nearest-neighbors to be $g_{1,2,3} = (2, -1, -1)$ for $d_{x^2-y^2}$ case, and $g_{1,2,3} = (0, 1, -1)$ for the $d_{xy}$ pairing state (as shown in Fig.~\ref{fig2}(g)). The result establishes that the $d+id$-pairing state in SLG at the VHS occurs between the NN sublattices with characteristic phases which accommodate the FS nesting features and corresponding sign-reversal in the gap structure.

\subsection{GBN}
In GBN, the circular FS allows small-angle nestings, and hence triplet pairing channel gains dominance, as in Sr$_2$RuO$_4$\cite{Sr2RuO4} and UPt$_3$\cite{UPt3}. This renders an odd-parity $p+ip$ wave pairing as shown in Fig.~\ref{fig2}(e). The symmetry belongs to the $E_{1}$ representation with two degenerate eigenfunctions\cite{KL_doped_graphene}:
\begin{eqnarray}
g^{p_x}({\bf k})&=&\sin\left(k_{y}-\phi_{\bf k}\right)+\sin\left(\frac{k_{y}}{2}+\phi_{\bf k}\right)\cos\left(\frac{\sqrt{3}k_{x}}{2}\right),\nonumber\\
g^{p_y}({\bf k})&=&\cos\left(\frac{k_{y}}{2}+\phi_{\bf k}\right)\sin\left(\frac{\sqrt{3}k_{x}}{2}\right).
\label{gkGBN}
\end{eqnarray}
Compared to the other two compounds, we find a considerably lower value of $\lambda=0.03$ in GBN. This is expected since this system does not have a strong nesting at a single wavevector, rather small-angle scattering wavevectors with lower strength. The inverse Fourier transformation of the pairing state yields $g_{1,2,3}=(2i, -i, -i)$ for the $p_{y}$ state and $g_{1,2,3}=(0, i, -i)$ for the $p_{x}$ state for the three NN Wannier sites (as shown in Fig.~\ref{fig2}(h)). Both $d+id$ - symmetry in SLG and $p+ip$ - wave pairing in GBN break time-reversal symmetry, and are chiral and nodeless in nature.

\subsection{TBG}
There have already been several proposals for unconventional pairing symmetries, and pairing mechanisms in TBG, such as  $d+id$\cite{TBGdSC,hexagonallattice1} as in SLG, odd-parity $p+ip$\cite{TBGpSC}, and others\cite{TBGotherSC}. The FS topology is quite different in TBG, exhibiting a three-fold symmetry for each valley. The three-fold symmetric FS is different from other triangular lattices with six-fold symmetric FS.\cite{NaCoO2} 
This FS topological change plays an important role in governing a distinct pairing symmetry in TBG. Here we obtain an extended $s$-wave pairing as shown in Fig.~\ref{fig2}(f), with its functional form given by
\begin{eqnarray}
g^{\rm ext-s}({\bf k})&=& 2\cos{(\sqrt{3}k_x/2)}\cos{\left(k_y/2\right)} + \cos{(k_y)}.
\label{gkTBG}
\end{eqnarray}
The pairing function is rotationally symmetric and changes sign between the \moire-zone center and corners, governing a symmetry that is consistent with the $A_{2g}$-group and hence called extended $s$-wave pairing. For the large FS at the VHS doping, the tip of the FS crosses through the nodal lines and thus gapless SC quasiparticle are obtained in this pairing state. This is a purely {\it real} gap function. In the direct \moire-lattice, this pairing symmetry stems from a nearest neighbor pairing between the Wannier sites in a triangular lattice given by $g_{1-6}=1$ for all components, see Fig.~\ref{fig2}(i).

We also note that the computed pairing symmetry in TBG is different from that of  the other triangular lattices, such as Na$_x$CoO$_2\cdot y$H$_2$O (NCOHO).\cite{NaCoO2} This is because the FS of NCOHO has the six-fold symmetry, while the FS for a given valley in TBG has three-fold symmetry.

\subsubsection{Valley dependent pairing symmetry in TBG}

\begin{figure}[h!]
\centering
\includegraphics[scale=0.19]{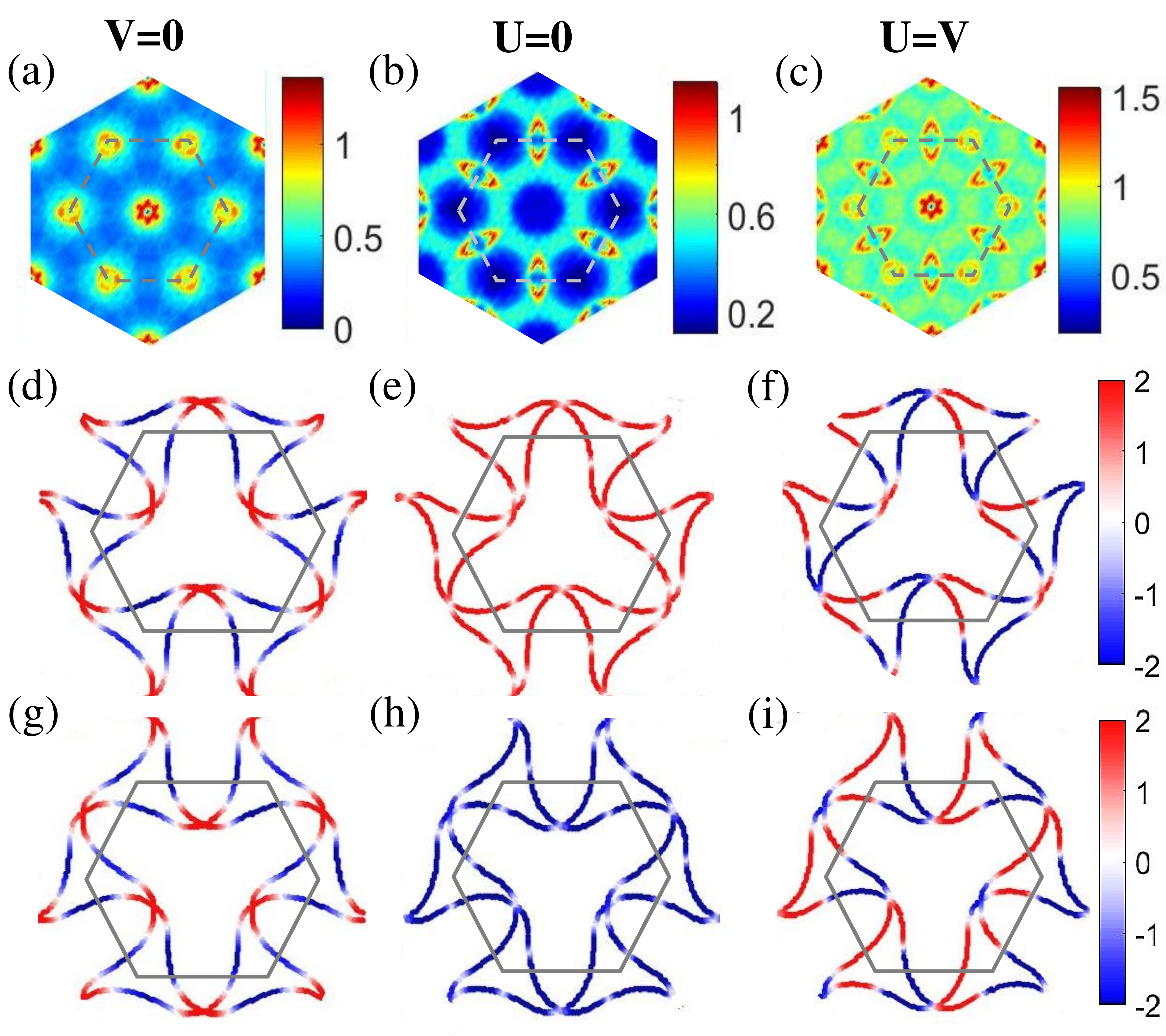}
\caption{(a-c) Spin susceptibility when (a) only intra-valley interaction ($U$) included with $V=0$, (b) only inter-valley interaction $V$ is included with $U=0$, and (c) both intra- and iter-valley interactions are included with $U=V$. (d-i) Computed pairing eigenfunctions for the highest eigenvalue of Eq. 4 in the main text for the corresponding cases in the upper panel. TBG at their VHS dopings are plotted on the FS in a blue (negative) to white (nodes) to red (positive) colormap. We separately plot the two valley result in different rows for easy visualization. (d-f) for one valley and (g-i) for the other valley.}
\label{pairing_TBG_valley}
\end{figure}

We repeat the calculation for the pairing eigenvalue and eigenfunctions by including both valley states for TBG. The FS for the two valleys are mutually rotated to each other by $\pi$. This changes the symmetry of the TBG lattice from triangular to hexagonal, as seen from the FS in Fig.~\ref{pairing_TBG_valley}(d-f). This opens up two competing nesting wavevectors $-$ intra-band and inter-band nestings $-$ as captured in the susceptibility result, see Fig.~\ref{pairing_TBG_valley}(a-c). We analyze the details of the pairing symmetry in the three limiting cases of (i) intra-valley interaction $U=3.5$ meV, inter-valley $V\rightarrow 0$, (ii) $U\rightarrow 0$, $V=3.5$ meV, and (iii) $U=V=3.5$ meV. In the three cases, we obtain extended $s$-, $s^{\pm}$- and $p+ip$-wave pairings, respectively. Below we discuss in details all three pairing states.

(i) First we consider the case for only intra-valley nesting in the limit of $U>>V$. Here the results are similar to the single-valley calculations shown in the main text. Consistently, we find an extended-$s$ wave symmetry for both valleys, where we have a sign reversal between center and corner of the BZ, with a circular nodal line (Fig.~\ref{pairing_TBG_valley}(d)). Inside the circle pairing value is positive and outside it is negative. We call it extended-$s$, because of the full rotational symmetry of the pairing function over the entire BZ.

(ii) Next we consider the case for only inter-valley nesting alone in the limit of $V>>U$. We obtain a completely different pairing symmetry. Here we find an {\it onsite}, $s$-wave pairing for each valley state, but the sign of the pairing is completely reversed between the two valleys, and hence called $s^{\pm}$-pairing state. The result is shown in Fig.~\ref{pairing_TBG_valley}(e). It is evident that the pairing symmetry does not have any $k$-dependence and arises solely from the onsite pairing of the Wannier orbitals, with different Wannier orbitals on the same site possess opposite phases. This pairing state is quite interesting in that while onsite pairing is often considered in the context of conventional, electron-phonon coupling cases, here one obtains an equivalent condition with an unconventional, electron-electron interaction, mechanism. Note that although the pairing interaction in obtained from the many-body electronic interaction, the strong onsite Coulomb repulsion potential is also present. Therefore, the onsite repulsion overturns the this onsite pairing strength, and such a onsite $s^{\pm}$ is disfavored.

(iii) Lastly we study the case of having both intra- and inter-valley nestings. The pairing eigenfunction map, plotted in Fig.~\ref{pairing_TBG_valley}(f) shows an approximate $p+ip$-pairing in a hexagonal lattice. We identify the pairing symmetry by identifying the corresponding nodal lines [see Fig.~\ref{pairing_TBG_valley}(d)] and by performing a reflection operation on any point of the FS. However, unlike previous cases, this symmetry contains higher harmonics of the $p$-wave symmetries as can be anticipated from complicated colormap of the pairing function on the FS. The pairing eigenvalue of this state is however much lower than the extended-$s$ wave pairing symmetry discussed above.

\subsubsection{Doping dependent pairing strength for TBG}
 
\begin{figure}[t]
\includegraphics[width=80mm,scale=1.0]{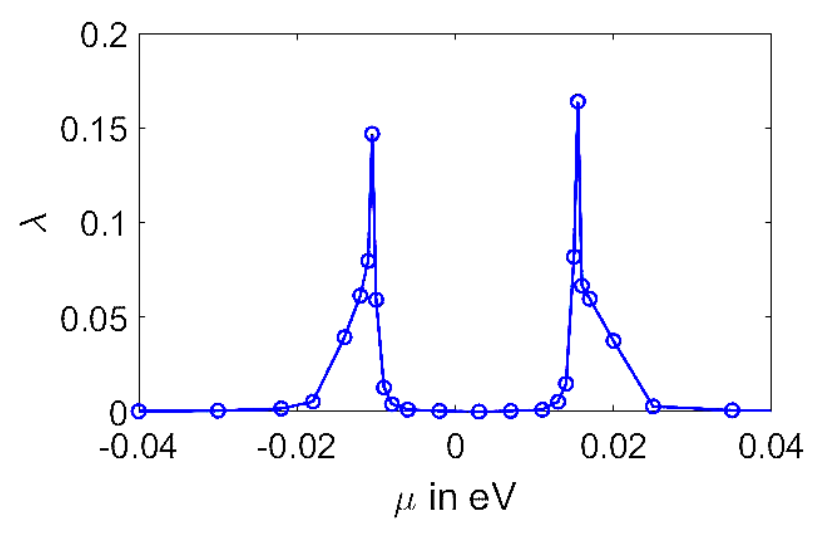}
\caption{Maximum pairing eigenvalue (SC coupling constant) $\lambda$ as a function of chemical potential shift $\mu$ for TBG. Note that the peaks in pairing eigenvalues occur when a flat band passes through the Fermi level.
}
\label{fig5}
\end{figure}

Finally, we study the doping dependence of the pairing eigenvalue $\lambda$, the SC coupling constant, in TBG, and the result is shown in Fig.~\ref{fig5}. We find that $\lambda$ attains maxima at the positions of the maxima of the density of states of the flatbands (roughly at half-fillings in both electron and hole doped sides). The present calculation does not include a correlated Mott gap. Mott gap opposes superconductivity and this will shift the SC maximum away from the half-filling, and one will reproduce the experimental phase diagram (work in progress).  

\section{Discussions and conclusions}\label{Sec:Discussion}
All the complex $d+id$, and $p+ip$ pairing symmetries do not possess SC gap nodes on the FS and thus their detection usually requires phase sensitive measurements. The extended $s$-wave one in TBG possess accidental nodes on both sides of the saddle-point near the VHS doping, and thus the SC gap is very anisotropic. The $k$-space mapping of the pairing symmetry can be measured via various modern techniques, such as angle-dependent photoemission spectroscopy, scanning tunneling probes via quasiparticle interference (QPI) pattern, field-angle dependence study of thermal conductivity, and so on. The nodal SC quasiparticle also leads to a power-law temperature dependence in many thermodynamical and transport properties which makes it easier to distinguish from conventional pairing. The sign reversal of the pairing symmetry leads to a magnetic spin-resonance at energy $<2\Delta$ ($\Delta$ is SC gap amplitude),\cite{magresonance} magnetic field dependence of QPI peaks,\cite{QPI} impurity resonance\cite{impurity} which all can be measured in future experiments for the verification of the underlying pairing symmetry.

In a typical unconventional superconductor, the Wannier states of the Fermi momenta are localized on each lattice site, and hence the correspondence between the reciprocal and direct lattice pairing is trivial. In the \moire~lattice, the location of the Wannier states corresponding to the flat band in TBG depends on energy, twist angle, and inter-layer coupling. In GBN, the Wannier states are localized on a hexagonal lattice. In TBG, they form a triangular lattice for each valley, where the hexagonal symmetry is restored when both valleys are included. Because of these materials specific peculiarities, the pairing symmetry of these materials turn out to be characteristically unique. The present paper spares several open questions for future studies. Superconductivity appears at a considerably low-carrier density ($\sim10^{12}$cm$^{-2}$), which may require adjustments in the theory. The competition between superconductivity and the correlated insulator gap is another interesting theme of research which will be perused in the future.

\begin{acknowledgments} 
We thank Priyo Adhikary for useful discussions. The work is supported by the Science and Engineering Research Board (SERB) of the Department of Science \& Technology (DST), Govt. of India for the Start Up Research Grant (Young Scientist), and also benefited from the financial support from the Infosys Science foundation under Young investigator Award.
\end{acknowledgments}

\begin{appendix}
\section{Model Hamiltonians for different systems}\label{AppA}

\subsection{SLG}
We use a tight-binding (TB)  model for SLG for our calculation taking into account nearest neighbour (NN) and the next nearest neighbour (NNN) hoppings. We start by describing the graphene lattice in terms of sublattices A and B with three NN translation vectors connecting sublaticce A to three NN-sublattices B as ${\bf \delta_{1}}=(\frac{1}{2},\frac{\sqrt{3}}{2})a_{0}$, ${\bf \delta_{2}}=(\frac{1}{2},-\frac{\sqrt{3}}{2})a_{0}$, ${\bf \delta_{3}}=(-\frac{1}{2},0)a_{0}$ with $a_{0}$ denoting the carbon-carbon distance in graphene lattice. Six NNN traslation lattice vectors can be written as ${\bf a}_1=\pm({\bf \delta_{1}}-{\bf \delta_{2}})$, ${\bf a}_2=\pm({\bf \delta_{2}}-{\bf \delta_{3}})$, ${\bf a}_3=\pm({\bf \delta_{3}}-{\bf \delta_{1}})$. We can write the Hamiltonian as
\begin{equation}
H_{\rm SLG} = H_{\rm on-site} + H_{\rm NN} + H_{\rm NNN}
\end{equation}
where,
\begin{eqnarray}
&& H_{\rm on-site}=\sum_{i,\sigma}\epsilon_{a} a^{\dagger}_{i,\sigma}a_{i,\sigma} + \sum_{j,\sigma}\epsilon_{b} b^{\dagger}_{j,\sigma}b_{j,\sigma} \\
&& H_{\rm NN}=-t\sum_{\langle i,j\rangle,\sigma} \left( a^{\dagger}_{i,\sigma}b_{j,\sigma} + h.c. \right) \\
&& H_{\rm NNN}=-t'\sum_{\langle\langle i,j\rangle\rangle,\sigma} \left( a^{\dagger}_{i,\sigma}a_{j,\sigma} + b^{\dagger}_{i,\sigma}b_{j,\sigma} + h.c. \right)
\end{eqnarray}
with $\epsilon_{a}$ and $\epsilon_{b}$ are sublattice energies for sublattice A and B respectively, $t$ and $t'$ are nearest neighbour and next nearest neighbour hopping amplitude respectively, $a^{\dagger}$ and $b^{\dagger}$ are creation operators on sublattices A and B respectively. Next we Fourier transform the creation and anihilation operators to get the band dispersion as
\begin{eqnarray}
&& H_{\rm on-site}=\sum_{{\bf k},\sigma} \left( \epsilon_{a} a^{\dagger}_{{\bf k},\sigma}a_{{\bf k},\sigma} + \epsilon_{b} b^{\dagger}_{{\bf k},\sigma}b_{{\bf k},\sigma} \right) \\
&& H_{\rm NN}=\sum_{{\bf k},\sigma} \left( \epsilon_{\bf k}^{\rm NN} a^{\dagger}_{{\bf k},\sigma}b_{{\bf k},\sigma} + h.c. \right) \\
&& H_{\rm NNN}=\sum_{{\bf k},\sigma} \left( \epsilon_{\bf k}^{\rm NNN} a^{\dagger}_{{\bf k},\sigma}a_{{\bf k},\sigma} + h.c. \right)
\end{eqnarray}
with
\begin{eqnarray}
&& \epsilon_{\bf k}^{\rm NN}=-t\sum_{i=1,2,3}e^{i{\bf k}.{\bf \delta}_{i}} \\
&& \epsilon_{\bf k}^{\rm NNN}=-t\sum_{i,j(i\ne j)}e^{i{\bf k}.\left( {\bf \delta}_{i}-{\bf \delta}_{j}\right)} 
\end{eqnarray}
The model with more tight-binding parameters and their values is given in Ref.~\cite{JJWannier}.

\subsection{GBN}
We construct the low energy model for graphene on hBN by following Ref.~\cite{JJ}. We write the four-band model in terms of $2 \times 2$ blocks given by
\begin{eqnarray}
H_{\rm GBN} = \left[
\begin{array}{ c c }
H_{\rm BN}  & T_{\rm BN,SLG}\\
T_{\rm SLG,BN} & H_{\rm SLG}
\end{array} \right],
\end{eqnarray}
where $H_{\rm BN}$ and $H_{\rm SLG}$ are Hamiltonians for Boron Nitride and SLG layers, respectively. $T_{\rm SLG,BN}$, $T_{\rm BN,SLG}$ are corresponding tunneling matrices in sublattice basis. The effective simplified model for this case is obtained by integrating out the boron nitride orbitals as $H=H_{\rm SLG}-T_{\rm SLG,BN}H_{\rm BN}^{-1}T_{\rm BN,SLG}$. Now the sub lattice dependent terms in the Hamiltonian can be written as
\begin{eqnarray}
H_{ss\prime}=H_{ss\prime}^{0}+H_{ss\prime}^{\rm MB},
\end{eqnarray}
where $H_{ss\prime}^{0}$ is the Hamiltonian that describes Dirac cones and $H_{ss\prime}^{\rm MB}$ gives the Moir\'{e} band modulation as
\begin{eqnarray}
H_{ss^{\prime}}^{0}&=&H_{ss^{\prime}}^{0}({\bf k,G}=0)\delta_{\bf k,k\prime},\\
H_{ss^{\prime}}^{MB}&=&\sum\limits_{{\bf G}\neq 0}H_{ss^{\prime}}^{\rm MB}({\bf k,G})\delta_{\bf k\prime-k-G}.
\end{eqnarray}
All the terms of the effective Hamiltonian now can be determined by the following equations
\begin{eqnarray}
&&H_{0}=C_{0}e^{i\phi_{0}},\ \ \ H_{z}=C_{z}e^{i\phi_{z}},\\
&&H_{\rm AA}=H_{0}+H_{z},\ \ \ H_{BB}=H_{0}-H_{z},\\
&&H_{\rm AB,{\bf G}_{1}}=H_{AB,{\bf G}_{4}}^{*}=C_{\rm AB}e^{i(2\pi/3-\phi_{\rm AB})},\\
&&H_{\rm AB,{\bf G}_{3}}=H_{AB,{\bf G}_{2}}^{*}=C_{\rm AB}e^{-i\phi_{\rm AB}},\\
&&H_{\rm AB,{\bf G}_{5}}=H_{AB,{\bf G}_{6}}^{*}=C_{\rm AB}e^{i(-2\pi/3-\phi_{\rm AB})}.
\end{eqnarray}
In Ref.~\cite{JJ}, it is shown that this effective model can be completely specified by six numbers $C_{0}=-10.13$ meV, $\phi_{0}=86.53^{0}$, $C_{z}=-9.01$ meV, $\phi_{0}=8.43^{0}$, $C_{\rm AB}=-11.34$ meV, $\phi_{\rm AB}=19.60^{0}$.

\subsection{TBG}

\begin{figure}[h!]
\centering
\includegraphics[scale=0.4]{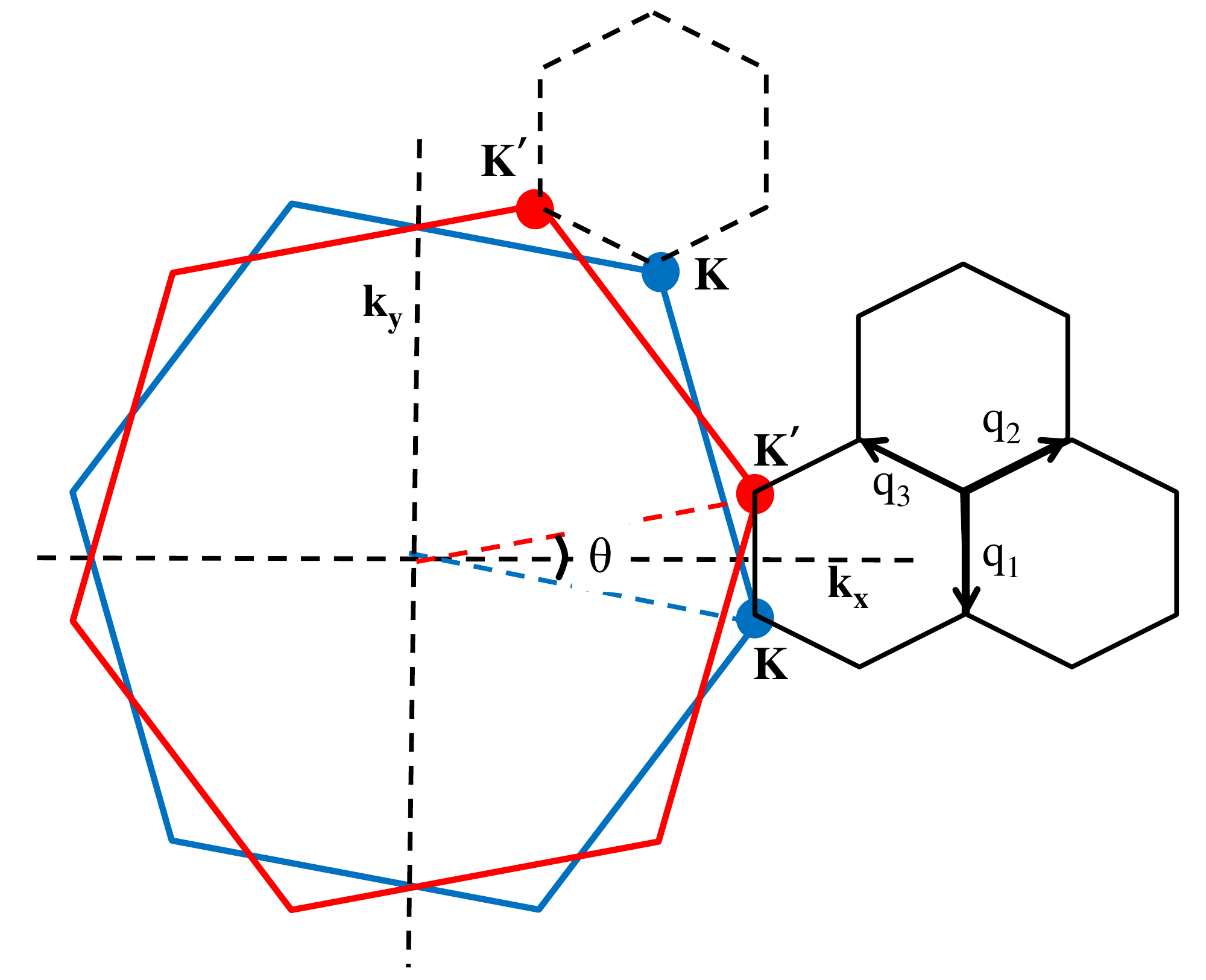}
\caption{Momentum-space formulation of TBG Moir\'{e} pattern. Red and blue BZ of SLG denotes the upper and lower layer, respectively. The upper layer is rotated by an angle $\theta/2$ and lower layer by $-\theta/2$ with respect to the $k_{x}$, $k_{y}$ axis shown in the figure. Smaller (solid black) hexagons represent the Moir\'{e} BZ of the TBG for a given valley state. Dashed black hexagon represents the Moir\'{e} BZ for the other valley state.}
\label{TBG}
\end{figure}

We construct the Hamiltonian for the TBG following the work of Bistritzer and MacDonald \cite{MacDonald}. We write down the low-energy Hamiltonian by considering two SLGs which were rotated by an angle $\theta$ with respect to each other and tunneling between the SLG layers (see Fig.~\ref{TBG}). Low-energy continuum model Hamiltonian for SLG can be written in a $2 \times 2$ matix as
\begin{eqnarray}
h_{\bf k}\left( \theta \right) = -vk\left[
\begin{array}{ c c }
0  & e^{i(\phi_{\bf k}-\theta)}\\
e^{i(\phi_{\bf k}-\theta)} & 0
\end{array} \right],
\end{eqnarray}
where $v$ = 3.2 eV\AA{}$^{-1}$ is the Dirac velocity, {\bf k} is the momentum measured from Dirac point, and $\phi_{\bf k}=\tan^{k_y/k_x}$, and $\theta$ is twist angle [see Fig.~\ref{TBG}]. Next we consider the inter-layer hoppings integrals, which can be accurately described by three distinct tunnelings with three distinct wavevectors ${\bf q}_{j}$ ($j=1,2,3$) [see Fig.~\ref{TBG}], whose directions are given by $(0,-1)$ for $j=1$, $(\sqrt{3}/2,1/2)$ for $j=2$, and $(-\sqrt{3}/2,1/2)$ for $j=3$.  The magnitude is $|{\bf q}_{j}|=2k_{D}\sin{(\theta/2)}$ where $k_{D}$ is the magnitude of BZ corner wavevector for a SLG. Corresponding tunneling matrices $T_{j}$ are given by
\begin{eqnarray}
T_{1} = c\left[
\begin{array}{ c c }
1  & 1\\
1  & 1
\end{array} \right], \ \ 
T_{2} =c \left[
\begin{array}{ c c }
e^{-i\zeta}  & 1\\
e^{i\zeta}  & e^{-i\zeta}
\end{array} \right], \ \ 
T_{3} = c\left[
\begin{array}{ c c }
e^{i\zeta}  & 1\\
e^{-i\zeta}  & e^{i\zeta}
\end{array} \right],\nonumber\\
\end{eqnarray}
where $\zeta=2\pi/3$. If the ${\bf k}$-cutoff is choosen in the first Moir\'{e} pattern BZ given by reciprocal lattice vectors $G_{1}=|{\bf q}_{j}|(\sqrt{3},0)$ and $G_{2}=|{\bf q}_{j}|(-\sqrt{3}/2,3/2)$. $c=0.9$ eV is the inter-layer tunneling amplitude. Now the Hamiltonian for TBG is a $8 \times 8$ matrix given by
\begin{eqnarray}
H_{\bf k} = \left[
\begin{array}{ c c c c }
h_{\bf k}\left( \theta/2 \right) & T_{1} & T_{2} & T_{3}\\
T_{1}^{\dagger} & h_{\bf q_{1}}\left( -\theta/2 \right) & 0 & 0\\
T_{2}^{\dagger} & 0 & h_{\bf q_{2}}\left( -\theta/2 \right) & 0\\
T_{3}^{\dagger} & 0 & 0 & h_{\bf q_{3}}\left( -\theta/2 \right)
\end{array} \right].\nonumber\\
\end{eqnarray}
We consider {\bf k}-points beyond the first shell approximation which resulted in a $400 \times 400$ matrix. After diagonalizing this matrix, we downfold the eigenvalues to the two (four) low-energy flat bands for a single valley (both valleys) that are near the FS, and all the subsequent calculations are performed considering only these bands.

\section{Calculation of pairing potential}\label{AppB}
We start with an extended Hubbard model both the valleys:
\begin{eqnarray}
\label{extended-Hubbard}
H_{\rm int}&=&\sum_{\alpha\beta,\sigma\sigma^{\prime},{\bf q}} U_{\alpha\beta}n_{\alpha\sigma}({\bf q})n_{\beta\sigma^{\prime}}(-{\bf q})\nonumber \\
 &=&U\sum_{\alpha,{\bf k,k^{\prime}},{\bf q}}c_{{\bf k}\alpha \uparrow}^{\dagger}c_{{\bf k+q}\alpha \uparrow}c_{{\bf k^{\prime}}\alpha \downarrow}^{\dagger}c_{{\bf k^{\prime}-q}\alpha \downarrow}\nonumber\\
&&+V\sum_{\alpha\ne\beta,{\bf k,k^{\prime},q},\sigma,\sigma^{\prime}}c_{{\bf k}\alpha \sigma}^{\dagger}c_{{\bf k+q}\alpha \sigma}c_{{\bf k^{\prime}}\beta\sigma^{\prime}}^{\dagger}c_{{\bf k^{\prime}-q}\beta\sigma^{\prime}},\nonumber\\
\end{eqnarray}
where $\alpha$ and $\beta$ are valley indices, taking values of  1 and 2 for two valleys in TBG. $c^{\dagger}$ and $c$ are creation and annihilation operators, respectively. $U$ and $V$ are intra-valley and inter-valley coupling strength respectively. In Eq.~\ref{extended-Hubbard} first term is the intra-valley interaction and second term is inter-valley interaction. By expanding Eq.~\ref{extended-Hubbard} to include multiple-interaction channels, we obtain the effective pairing potential $\Gamma_{\alpha\beta}({\bf k}-{\bf k^{\prime}})$  for the singlet and triplet states. The corresponding pairing Hamiltonian is 
\begin{eqnarray}
\label{superconducting}
H_{\rm int} \approx \sum_{\alpha\beta,{\bf k,k'},\sigma,\sigma'} \Gamma_{\alpha\beta}({\bf k}-{\bf k^{\prime}}) c_{{\bf k}\alpha \sigma}^{\dagger}c_{-{\bf k}\alpha \sigma^{\prime}}^{\dagger}c_{-{\bf k^{\prime}}\beta\sigma^{\prime}}c_{{\bf k^{\prime}}\beta\sigma}.\nonumber\\
\end{eqnarray}
The pairing potentials are
\begin{eqnarray}
\tilde{\Gamma}^{\rm s}_{\alpha\beta}({\bf q})= \frac{1}{2}{\rm Re}\big[3{\tilde U}^{s}{\tilde \chi}^{s}_{\alpha\beta}({\bf q}){\tilde U}^{s} - {\tilde U}^{c}{\tilde \chi}^{c}_{\alpha\beta}({\bf q}){\tilde U}^{c} + {\tilde U}^{s}+{\tilde U}^{c}\big],\nonumber\\
\label{singlet_valley}\\ \nonumber\\
\tilde{\Gamma}^{\rm t}_{\alpha\beta}({\bf q})= -\frac{1}{2}{\rm Re}\big[{\tilde U}^{s}{\tilde \chi}^{s}_{\alpha\beta}({\bf q}){\tilde U}^{s} + {\tilde U}^{c}{\tilde \chi}^{c}_{\alpha\beta}({\bf q}){\tilde U}^{c}- {\tilde U}^{s}-{\tilde U}^{c}\big].\nonumber\\
\\
\label{triplet_valley} \nonumber
\end{eqnarray}
Here $U^{s/c}=U$ for $\alpha=\beta$ and $U^{s/c}=V$ for $\alpha \ne \beta$. From the superconducting Hamiltonian Eq.~\ref{superconducting} we can construct the superconducting gap (SC) equation as
\begin{eqnarray}
\Delta_{n,{\bf k}}^{\alpha}&&= -\sum_{\beta,{\bf k^{\prime}}}\Gamma^{n}_{\alpha\beta}({\bf k}-{\bf k^{\prime}})\left\langle c_{-{\bf k^{\prime}}\beta\sigma}c_{{\bf k^{\prime}}\beta\sigma'} \right\rangle
\end{eqnarray}
Here $n=s,t$ for singlet and triplet pairing channels where $\sigma'=\mp\sigma$, respetively. In the limit $T \rightarrow 0$ we have $\left\langle c_{-{\bf k^{\prime}}\beta\sigma}c_{{\bf k^{\prime}}\beta\sigma'}\right\rangle \rightarrow \lambda_n\Delta_{n,{\bf k^{\prime}}}^{\beta}$ which makes the above equation an eigenvalue equation
\begin{eqnarray}
\Delta_{n{\bf k}}^{\alpha}&&= -\lambda\sum_{\beta,{\bf k^{\prime}}}\Gamma^n_{\alpha\beta}({\bf k}-{\bf k^{\prime}})\Delta_{n,{\bf k^{\prime}}}^{\beta}.
\end{eqnarray}
In our work we solve the eigenvalue problem separately for the singlet and triplet channels. The following equations remain the same for both these pairing channels and thus the index '$n$' is omitted for simplicity. This is an eigenvalue equation for the k-points in the Fermi surface $(\Delta_{\bf k_{F}}^{\alpha})$. For this purpose we construct the matrix
\begin{eqnarray}
\Gamma({\bf k_{F}}-{\bf k^{\prime}_{F}}) = \left[
\begin{array}{ c c c}
\Gamma_{{\bf k_{F}}{\bf k_{F}}^{\prime}}^{11}  & \Gamma_{{\bf k_{F}}{\bf k_{F}}^{\prime}}^{12} & \dots\\
\Gamma_{{\bf k_{F}}{\bf k_{F}}^{\prime}}^{21}  & \Gamma_{{\bf k_{F}}{\bf k_{F}}^{\prime}}^{22} & \dots\\
\vdots & \vdots & \ddots
\end{array} \right],
\end{eqnarray}
where 1,2, refer to the band/valley indices, and ${\bf q}={\bf k_{F}}-{\bf k^{\prime}_{F}}$ the Fermi surface nesting vctor and $\Gamma_{{\bf k_{F}}{\bf k_{F}}^{\prime}}^{\alpha\beta}$ refers to $N\times N$ matrix if $N$ number of points on the Fermi surface is considered for each valley. Now if we denote ${\bf \Delta_{k_{F}}}=\left[ \Delta_{\bf k_{F}}^{1} \ \ \ \Delta_{\bf k_{F}}^{2} \right]^{T}$ then we can write the matrix equation and solve for its eigenvalues and eigenvectors as
\begin{eqnarray}
{\bf \Delta_{k_{F}}}=-\lambda\sum_{{\bf k}'_{\rm F}} {\bf \Gamma(k_{F}-k^{\prime}_{F})\Delta_{k_{F}^{\prime}}}.
\end{eqnarray}
By writing the SC gap function as $\Delta_{\bf k}=\Delta_0g_{\bf k}$, where $\Delta_0$ is the gap amplitude and $g_{\bf k}$ is the gap anisotropy, we obtain Eq. 4 in the main text.

\end{appendix}


\begin{thebibliography}{10}
\bibitem{TB_technicaldetails}For GBN, we use the full TB dispersion presented in Ref.~\onlinecite{JJ}. For TBG we calculate band structure using continuum model where interlayer hopping is allowed for three distinct wave vectors whose directions are given by $(0,-1)$, $(-\sqrt{3}/2,1/2)$ and $(\sqrt{3}/2,1/2)$. We diagonalize a 392$\times$392 matrix extending the cutoff wave vector outside the first Brillouin zone.
\bibitem{U}The onsite Hubbard interaction becomes materials dependent due to varying screening effects, as evident in Fig.~\ref{fig2}, and also shown in Ref.~\onlinecite{ADas}. We use materials dependent Hubbard $U$ (same for all bands): $U=$1~eV for SLG, 100~meV for GBN, and 3~meV for TBG. For TBG the insulating gap is $\sim 0.3$meV$<U$. This can be reconciled for a dispersive Hubbard band within the Mott picture or for the Slater picture where the gap is $U$ times order parameter.
\bibitem{NNdistances}For SLG, the three nearest-neighbor (NN) distances for the pairings are\cite{CastroNetoRMP} ${\bm \delta}_{1,2}=(1,\pm\sqrt{3})a/2$, and ${\bm \delta}_{3}=(-1,0)a$, where $a=2.46\AA$ is the lattice constant. For GBN, the next nearest neighbor (NNN) sites are involved in pairing with odd-parity (as shown in Fig.~\ref{fig2}), where the NNN positions (on both sides) are ${\bm \delta}^{\prime}_{1,2}=(3,\pm\sqrt{3})a'/2$, and ${\bm \delta}^{\prime}_{3}=(0,-\sqrt{3})a'$, with $a'\sim 40 a$ is the \moire-lattice constant. For TBG, the triangular lattice sites have the NN distances (on both sides) as: ${\bm \delta}_{1}=(-1,0)a'$, ${\bm \delta}_{2,3}=(\pm 1,\sqrt{3})a'/2$, where the corresponding \moire-lattice constant $a'\sim 40a$ near the magic angles
\end{thebibliography}
\end{document}